\documentclass[12pt]{article}
\usepackage{geometry}
\usepackage{makeidx}
\usepackage[T1]{fontenc}
\usepackage[dvipsnames,svgnames,table]{xcolor}
\usepackage{epstopdf}
\usepackage{epsfig}
\usepackage{textcomp}
\usepackage{hyperref}
\usepackage{amsmath}
\usepackage{amssymb}
\usepackage{multirow}
\usepackage{multicol}
\usepackage{booktabs}
\usepackage{lastpage}
\usepackage{hyperref} 
\usepackage{graphicx}
\usepackage{enumerate}
\usepackage{threeparttable}
\usepackage{float}
\usepackage{array}
\usepackage{cite}
\usepackage{color,soul}
\usepackage{listings}
\usepackage{tikz}
\makeatletter
\renewcommand\section{\@startsection{section}{1}{\z@}%
                    {-2.5ex \@plus -1ex \@minus -.2ex}%
                    {2.3ex \@plus.2ex}%
                    {\normalfont\large\bfseries}}
\renewcommand\subsection{\@startsection{subsection}{1}{\z@}%
                    {-2.5ex \@plus -1ex \@minus -.2ex}%
                    {2.3ex \@plus.2ex}%
                    {\small\bfseries}}

\begin{document}
\bigskip
\title{\textbf{Implementation of atmospheric proton spectrum in GEANT4 simulations for space applications}}
\medskip
\author{\small A. Ilker Topuz$^{1}$}
\medskip
\date{\small$^1$Institute of Physics, University of Tartu, W. Ostwaldi 1, 50411, Tartu, Estonia}
\maketitle
\begin{abstract}
A major part of cosmic rays consists of the primary protons, and this portion plays a crucial role in the space applications such as shielding of spacecrafts. In this study, the proton flux values measured at the top of the atmosphere through the BESS-TeV spectrometer in 2004 are introduced into the GEANT4 simulations by using a probability grid that generates the corresponding discrete kinetic energies with a certain discrete probability. The introduced scheme is tested over a set of the shielding materials such as aluminum, polypropylene, Kevlar, polyethylene, and water by computing the total absorbed dose, which is the measure of the cumulative energy deposited in the investigated target volumes by protons per unit mass in Gy. It is shown that the present recipe provides the opportunity to use the discrete energy values together with the experimental flux values, thereby demonstrating a beneficial capability in the GEANT4 simulations for diverse space applications.
\end{abstract}
\section{Introduction}
While the primary composition of cosmic rays includes different particles, the primary cosmic ray protons constitute the dominant part together with the alpha particles~\cite{buchvarova2022galactic, bowman2022investigating}. Among the notable experimental studies that were conducted for the measurement of the cosmic ray particles is the BESS-TeV spectrometer in 2004 where the tabulated proton flux values are provided in [m$^{-2}$s$^{-1}$sr$^{-1}$GeV$^{-1}$] at the discrete energies starting from 1.08 GeV up to 463 GeV~\cite{haino2004measurements}.

By reminding that the major portion of the cosmic rays consists of the primary protons, the energy spectrum as well as the corresponding population of these primary particles becomes an important aspect in the development of shielding materials for the space applications. Motivated by the implementation of the experimental energy spectrum for the primary cosmic ray protons with the aim of the space applications in the GEANT4 simulations~\cite{agostinelli2003geant4}, the proton energy spectrum measured by the BESS-TeV spectrometer in 2004 is introduced into GEANT4 by using a probability grid that converts the proton flux values into the discrete probabilities. Following the implementation in the GEANT4 code, the primary cosmic proton spectrum is checked over a number of spacecraft materials including aluminum, polypropylene, Kevlar, polyethylene, and water~\cite{finckenor2018materials} by calculating the cumulative absorbed dose. 

This study is organized as follows. In section~\ref{Implementation}, the implementation of the discrete probabilities at the discrete energies via the probability grid is shown in addition to the proton flux variation through the BESS-TeV spectrometer for an energy interval lying between 0 and 68 GeV. While the proton spectrum is tested over the shielding materials by computing the total absorbed dose in section~\ref{Application}, the conclusions are drawn in section~\ref{Conclusion}.
\section{Implementation in GEANT4}
\label{Implementation}
\begin{figure}[H]
\begin{center}
\includegraphics[width=13cm]{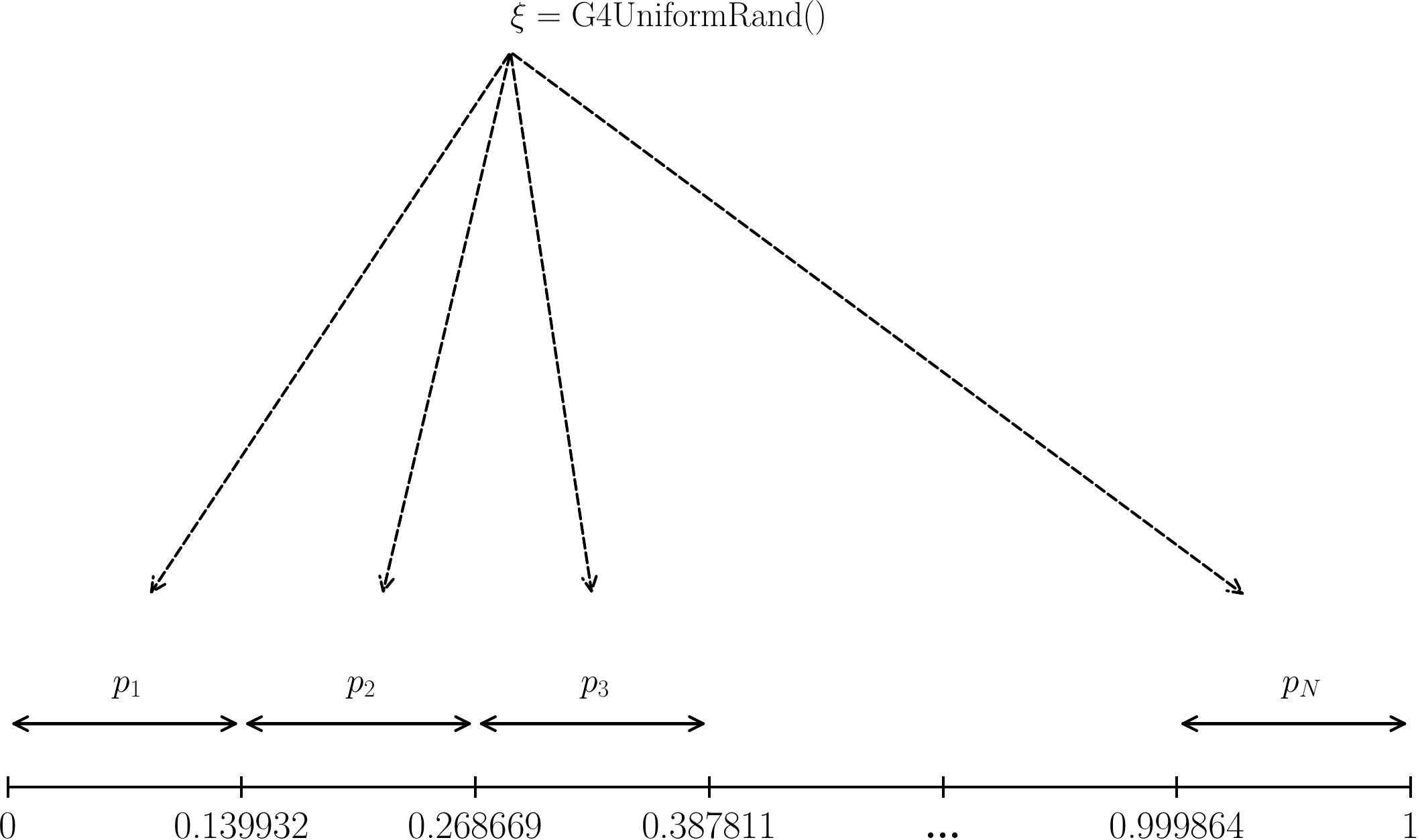}
\caption{Probability grid that assigns the discrete kinetic energies via the discrete probabilities in accordance with Table~\ref{BESSlist}.}
\label{grid}
\end{center}
\end{figure}

The kinetic energies of the primary protons denoted by $E_{i}$ as well as the corresponding flux values labeled as $\phi_{p,i}$ at the top of the atmosphere are acquired from another study that is dedicated to the measurement of the primary and atmospheric cosmic-ray spectra with the BESS-TeV spectrometer in 2004~\cite{haino2004measurements} as listed in Table~\ref{BESSlist}. 
\begin{table}[H]
\begin{center}
\caption{BESS-TeV discrete experimental flux values and discrete probabilities at the corresponding discrete energies between 0 and 68 GeV.}
\begin{tabular*}{0.48\columnwidth}{@{\extracolsep{\fill}}*2c}
\toprule
\toprule
$E_{i}$ [GeV] & $\phi_{p,i}$ [m$^{-2}$s$^{-1}$sr$^{-1}$GeV$^{-1}$]\\
\midrule
0.00&0.000\\
1.08&350.000\\
1.26&322.000\\
1.47&298.000\\
1.71&271.000\\
2.00&241.000\\
2.33&208.000\\
2.71&174.000\\
3.16&145.000\\
3.69&119.000\\
4.30&95.200\\
5.01&73.500\\
5.84&56.300\\
6.81&43.400\\
7.93&31.500\\
9.25&22.500\\
10.80&15.900\\
12.60&11.200\\
14.70&7.710\\
17.10&5.330\\
19.90&3.630\\
23.20&2.480\\
27.10&1.620\\
31.60&1.090\\
36.80&0.717\\	
42.90&0.484\\
50.00&0.315\\
58.30&0.207\\
68.00&0.134\\
\bottomrule
\bottomrule
\end{tabular*}
\begin{tabular*}{0.48\columnwidth}{@{\extracolsep{\fill}}*2c}
\toprule
\toprule
$E_{i}$ [GeV] & $p_{i}$\\
\midrule
0.00&0\\
1.08&0.139931881\\
1.26&0.128737331\\
1.47&0.119142002\\
1.71&0.108347257\\
2.00&0.096353095\\
2.33&0.083159518\\
2.71&0.069566135\\
3.16&0.057971779\\
3.69&0.047576840\\
4.30&0.038061472\\
5.01&0.029385695\\
5.84&0.022509043\\
6.81&0.017351553\\
7.93&0.012593869\\
9.25&0.008995621\\
10.80&0.006356905\\
12.60&0.004477820\\
14.70&0.003082499\\
17.10&0.002130963\\
19.90&0.001451294\\
23.20&0.000991517\\
27.10&0.000647685\\
31.60&0.000435788\\
36.80&0.000286660\\
42.90&0.000193506\\
50.00&0.000125939\\
58.30&0.000082759\\
68.00&0.000053573\\
\bottomrule
\bottomrule
\end{tabular*}
\label{BESSlist}
\end{center}
\end{table}
First, the total proton flux between 0 and 68 GeV is computed by adding up each flux value at a given kinetic energy over 28 bins, and the probability for a particular kinetic energy is calculated through the ratio between the corresponding flux value at this specific kinetic energy and the total proton flux. Then, the discrete probabilities denoted by $p_{i}$ for the energy interval lying between 0 and 68 GeV are determined as written in
\begin{equation}
p_{i}=\frac{\phi_{p,i}}{\sum\limits_{i=0}^{28}\phi_{p,i}}~~~{\rm with}~~~\sum\limits_{i=0}^{28}p_{i}=1
\label{BESSprob}
\end{equation}
As shown in Table~\ref{BESSlist}, the proton energy spectrum is a fast decaying function, the values of which are diminished by three orders of magnitude when the kinetic energy is 68 GeV. 
\begin{figure}[H]
\begin{center}
\includegraphics[width=8cm]{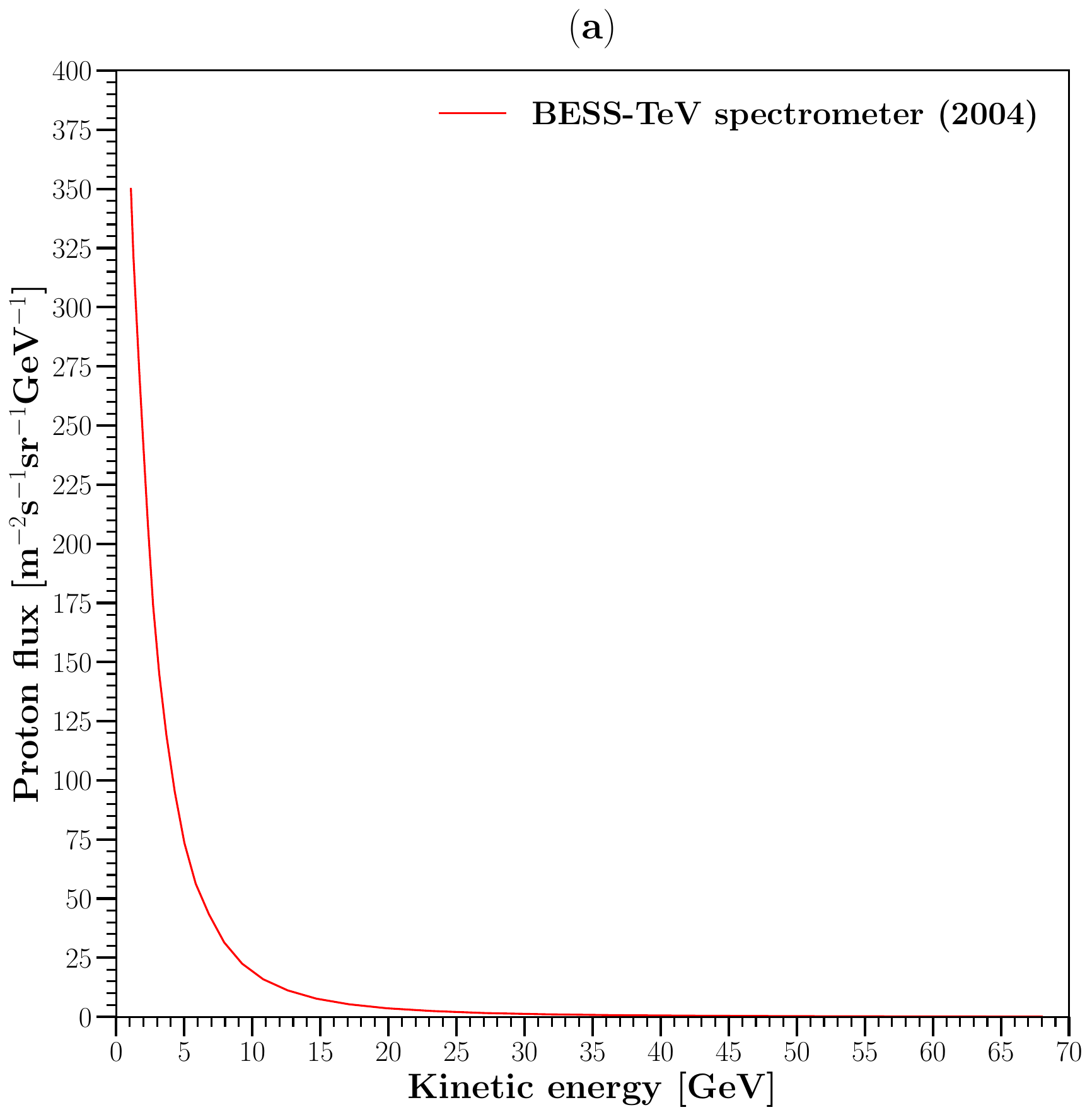}
\includegraphics[width=8cm]{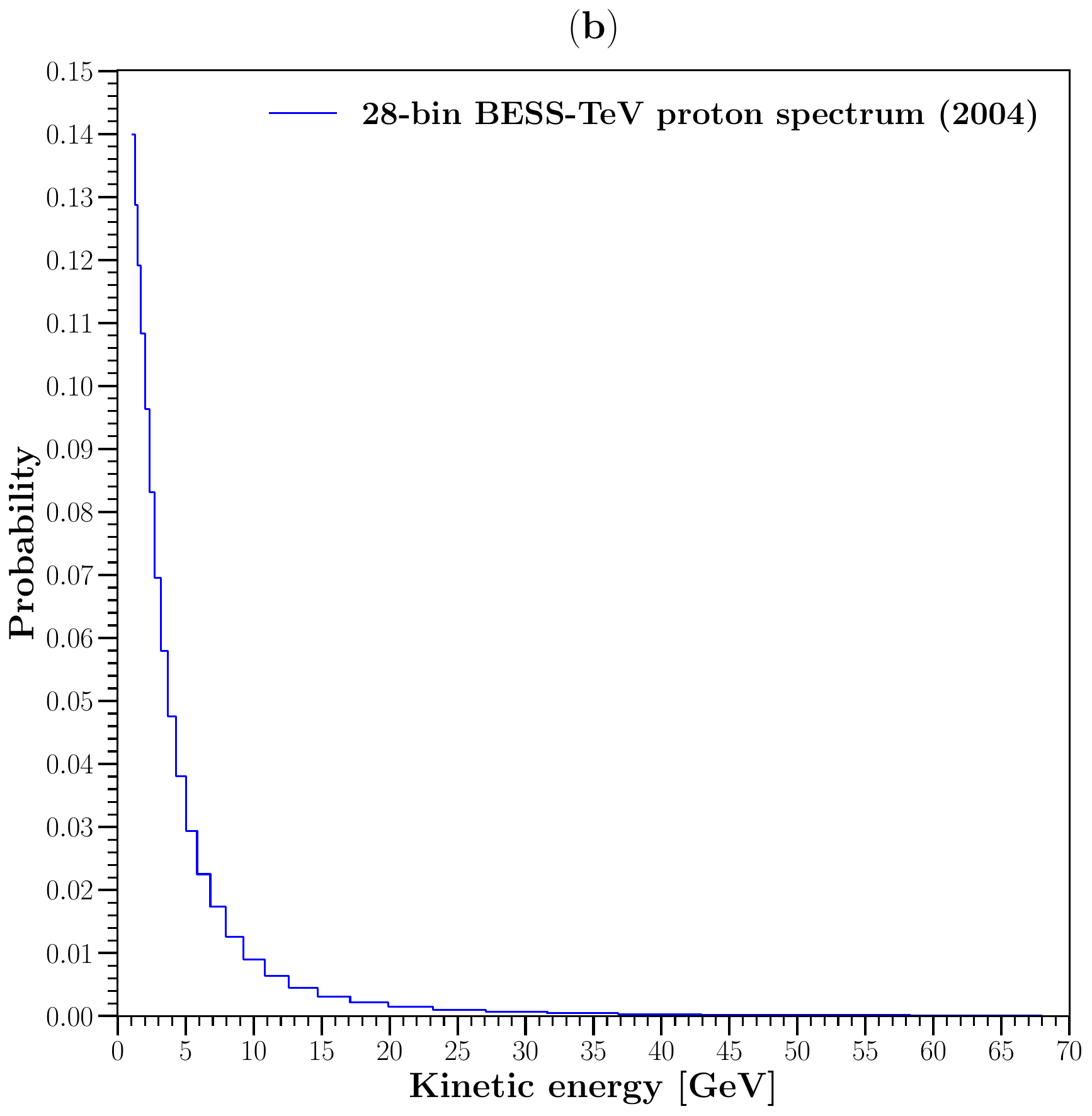}
\includegraphics[width=8cm]{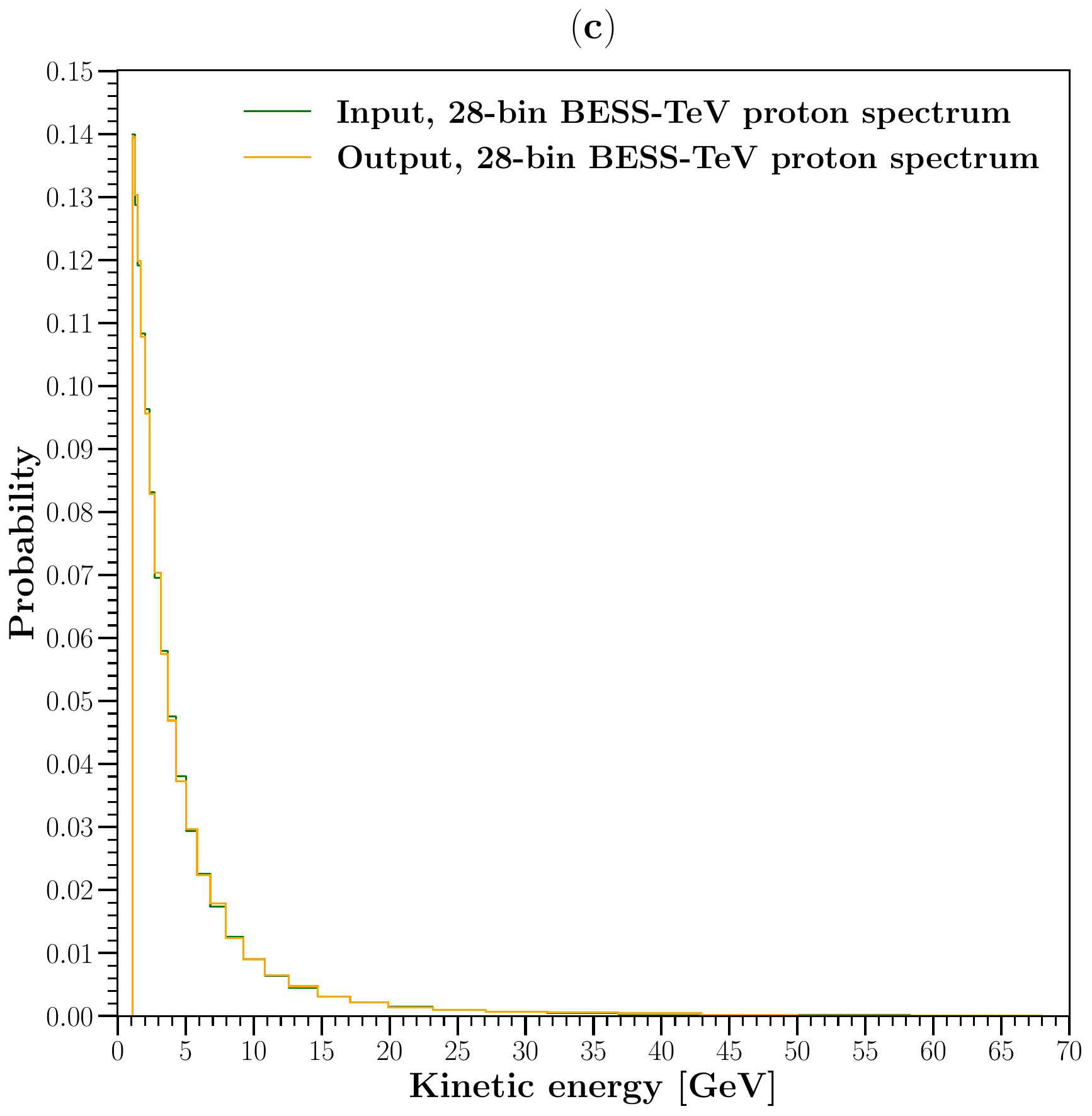}
\caption{BESS-TeV proton flux values at the top of the atmosphere~\cite{haino2004measurements}: (a) Variation of the experimental flux between 0 and 68 GeV, (b) Histogram of discrete probabilities between 0 and 68 GeV, (c) Contrast between the input dataset before processing with the probability grid and the output values through the activation of the present probability grid.}
\end{center}
\label{fluxvar}
\end{figure}
The variation of the experimental proton flux is shown in Fig.~\ref{fluxvar}(a), while the histogram of the discrete probabilities for the kinetic energies between 0 and 68 GeV is depicted in Fig.~\ref{fluxvar}(b). The average kinetic energy for the cosmic protons that exist at the top of the atmosphere is around 2.69 GeV for a kinetic energy interval between 0 and 68 GeV according to the experimental measurements via the BESS-TeV spectrometer in 2004. As illustrated in Fig.~\ref{fluxvar}(a) and (b), the maximum probability occurs at 1.08 GeV, whereas 68 GeV indicates the minimum probability in the present study.
For the purpose of implementing the discrete proton energy spectrum, the present strategy to inject the incoming protons is subsequently integrated by means of G4ParticleGun as can be found in the previous studies~\cite{topuz2022towards, topuz2022dome, topuz2023particle, topuz2023come}. By recalling the unity condition, a grid is built by summing up the discrete probabilities, the interval of which starts with 0 and ends in 1 as illustrated in Fig.~\ref{grid}. Thus, each cell in this grid, i.e. the difference between two points on the probability grid, specifies a discrete probability. 
Then, a random number denoted by $\xi$ between 0 and 1 is generated by using the pre-defined uniform number generator called G4UniformRand(). Finally, this random number is scanned on the probability grid by checking the difference between the grid points, and the particular discrete energy is assigned when the random number matches with the associated cell. This operation is summarized as explained in
\begin{equation}
{\rm If}~\sum\limits_{i=0}^{x}p_{i}<\xi\leq \sum\limits_{i=0}^{x+1}p_{i} {\rm~for~any}~x\in\{0,1,2,...,N-1\}, {\rm~then~Energy}=E_{x+1}
\end{equation}
In the wake of the above-mentioned implementation, the discrete probabilities listed in Table~\ref{BESSlist} that are introduced in the GEANT4 code are verified, and the comparison between Table~\ref{BESSlist} and the code output is demonstrated in Fig.~\ref{fluxvar}(c). 
\section{Application for space shielding materials}
\label{Application}
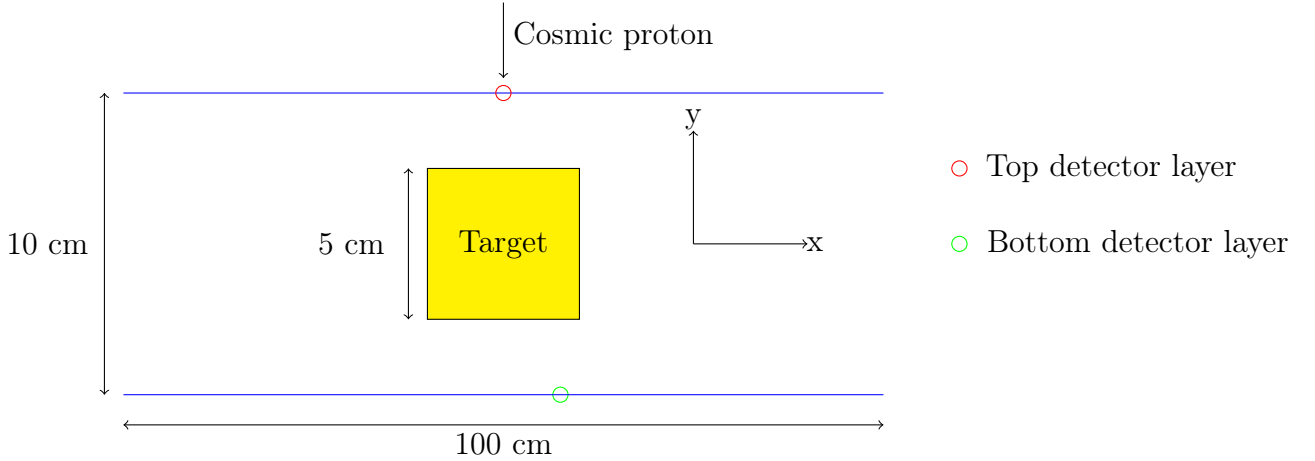
\begin{figure}[H]
\begin{center}
\begin{tikzpicture}
\draw[->] (2.5,0) -- (4,0);
\node at (4.1,0) {x};
\draw[->] (2.5,0) -- (2.5,1.5);
\node at (2.5,1.65) {y};
\node at (1.45,2.75) {Cosmic proton};
\draw[->] (0,3.2) -- (0,2.2);
\draw [blue] (-5,2) -- (5,2);
\draw[red] (0,2) circle (0.1cm);
\draw[green] (0.75,-2) circle (0.1cm);
\draw[red] (6,1) circle (0.1cm);
\node at (8,1.0) {Top detector layer};
\draw[green] (6,0) circle (0.1cm);
\node at (8.35,0.0) {Bottom detector layer};
\draw (0,0) node[minimum size=2cm,fill=yellow,draw] {\mbox{Target}};
\draw [blue] (-5,-2) -- (5,-2);
\draw[<->] (5,-2.4) -- (-5,-2.4);
\node at (0,-2.65) {100 cm};
\draw[<->] (-5.25,-2) -- (-5.25,2);
\node at (-6,0) {10 cm};
\draw[<->] (-1.25,-1) -- (-1.25,1);
\node at (-2,0) {5 cm};
\end{tikzpicture}
\caption{Simulation setup consisting of top detector layer as well as bottom detector layer to compute the energy loss.}
\label{detectorsetup}
\end{center}
\end{figure}

The 28-bin discrete cosmic proton spectrum is tested on a list of shielding materials that is composed of aluminum, polypropylene, Kevlar, polyethylene, and water over a detector setup as depicted in Fig.~\ref{detectorsetup}. The kinetic energies of the incoming protons are collected from the top detector layer as well as the bottom detector layer in order to calculate the energy loss due to the target material. The dimensions of the detector layers are $100\times0.4\times100$ cm$^{3}$, while the cubic volume of the shielding materials is 125 cm$^{3}$. The distance between the detector layers is 10 cm, and the target material is situated at the center of the detector configuration.
Through the detector setup demonstrated in Fig.~\ref{detectorsetup}, the total absorbed dose due to the primary cosmic protons at the top of the atmosphere is determined by using the ratio of the kinetic energy difference between the detector sections over the mass of the target material as follows
\begin{equation}
D_{\rm Primary}=\sum\Delta E_{\rm Kinetic}^{\rm Primary}/{\rm m}
\end{equation}
where $D_{\rm Primary}$ is the total absorbed dose due to only the primary protons, $\Delta E_{\rm Kinetic}$ is the kinetic energy difference between the detector sections, and $m$ is the mass of the target volume. The total absorbed dose in MeV/kg is converted into Gy (J/kg) by utilizing a coefficient of $1.60217663\times10^{-13}$. 

A series of GEANT4 simulations are performed to compute the total energy loss as well as the total absorbed dose by using the 28-bin discrete cosmic proton spectrum, and the simulations properties are tabulated in Table~\ref{Simulationproperties}.  A narrow planar multi-energetic mono-directional beam that is generated at ([-0.5, 0.5] cm, 85 cm, [-0.5, 0.5] cm) via G4ParticleGun is utilized, and the generated protons are propagating in the vertically downward direction as shown by the black arrow in Fig.~\ref{detectorsetup}, i.e. from the top edge of the simulation box through the bottom edge. The total number of the generated cosmic protons is $5\times10^{4}$ in every simulation. All the materials in the simulation geometry are defined in agreement with the GEANT4/NIST material database, and QBBC is the reference physics list used in the present study.
\begin{table}[H]
\begin{center}
\caption{Simulation properties.}
\begin{tabular}{cc}
\toprule
\toprule
Particle & proton\\
Beam direction & Vertical\\
Momentum direction & (0, -1, 0)\\
Source geometry & Planar\\
Initial position (cm) & ([-0.5, 0.5], 85, [-0.5, 0.5])\\
Number of particles & $5\times10^{4}$\\
Energy interval (GeV) & [0, 68]\\
Energy cut-off (GeV) & 1.08\\
Average kinetic energy (GeV) & 2.69\\
Energy distribution & BESS-TeV\\
Target geometry & Cube\\
Target volume (cm$^{3}$) & 5$\times$5$\times$5\\ 
Material database & G4/NIST\\
Reference physics list & QBBC\\
\bottomrule
\bottomrule
\label{Simulationproperties}
\end{tabular}
\end{center}
\end{table}
The proton tracking is maintained by G4Step, and the registered energy values are post-processed by the aid of a Python script where the energy difference is first calculated for every single primary proton exiting from the shielding material, then the energy loss of each detected primary proton is divided by the target mass and converted to Gy. Only the primary protons that hit the two detector layers are taken into consideration, and the secondary particles generated due to the inelastic interactions are neglected.

Regarding the simulations outcomes, the calculated total energy loss as well as the computed total absorbed doses for aluminum, polypropylene, Kevlar, polyethylene, and water are listed in Table~\ref{Simulationoutcomes}.
\begin{table}[H]
\begin{center}
\caption{Simulation outcomes over $5\times10^{4}$ cosmic primary protons.}
\begin{tabular*}{\columnwidth}{@{\extracolsep{\fill}}*3c}
\toprule
\toprule
Material & $\sum\Delta E_{\rm Kinetic}^{\rm Primary}$ [MeV] & $D_{\rm Primary}$ [Gy]\\
\midrule
Aluminum & $1.011\times10^{6}$ & $0.479\times10^{-6}$\\
Polypropylene & $0.536\times10^{6}$ & $0.764\times10^{-6}$\\ 
Kevlar & $0.694\times10^{6}$ & $0.619\times10^{-6}$\\
Polyethylene& $0.549\times10^{6}$ & $0.749\times10^{-6}$ \\
Water& $0.545\times10^{6}$ & $0.698\times10^{-6}$ \\
\bottomrule
\bottomrule
\label{Simulationoutcomes}
\end{tabular*}
\end{center}
\end{table}
It is observed that the maximum total energy deposition occurs in the case of aluminum, whereas polypropylene yields the maximum total absorbed dose. Contrary to this trend, the minimum total energy loss exists in the case of polypropylene, while the minimum total absorbed dose is obtained in the simulations of aluminum. Both polypropylene and polyethylene exhibit the same dosimetric profile.

In order to reveal out the dosimetric contribution of the all the secondary particles, the total energy deposited within the target volume is computed by including the primary protons as well as the secondaries generated via $5\times10^{4}$ initial cosmic primary protons.  Hence, the total absorbed dose denoted by $D_{\rm Total}$ is calculated by using the following expression:
\begin{equation}
D_{\rm Total}=\sum E_{\rm Deposited}/{\rm m}
\end{equation}
where $E_{\rm Deposited}$ is the energy deposited by all the generated particles crossing the target volume. In contrast with the previous procedure, $\sum E_{\rm Deposited}$ is determined by registering the total energy deposited due to all the particles within the target volume, which is defined via G4Step without using any detector component. The simulation results are tabulated in Table~\ref{Simulationoutcomes2}.
\begin{table}[H]
\begin{center}
\caption{Simulation outcomes including all the secondaries within the target volume.}
\begin{tabular*}{\columnwidth}{@{\extracolsep{\fill}}*3c}
\toprule
\toprule
Material & $\sum E_{\rm Deposited}$ [MeV] & $D_{\rm Total}$ [Gy]\\
\midrule
Aluminum & $1.915\times10^{6}$ & 0.909$\times10^{-6}$\\
Polypropylene & $0.654\times10^{6}$ & 0.931$\times10^{-6}$\\ 
Kevlar & $1.024\times10^{6}$ & 0.911$\times10^{-6}$\\
Polyethylene & $0.686\times10^{6}$ & 0.936$\times10^{-6}$ \\
Water& $0.692\times10^{6}$ & 0.887$\times10^{-6}$ \\
\bottomrule
\bottomrule
\label{Simulationoutcomes2}
\end{tabular*}
\end{center}
\end{table}
From the simulation outcomes in Table~\ref{Simulationoutcomes2}, it is shown that the total energy deposited within the target material is significantly augmented if the secondary particles are taken into account. Especially in the case of aluminum, the secondary particles have a significant contribution to the total energy deposited and consequently to the total absorbed dose. Moreover, both polypropylene and polyethylene lead again to the similar dosimetric values. 
\section{Conclusion}
The present recipe might be beneficial in the cases where the experimental flux values along with the discrete energy values are provided. It is seen that a discrete proton spectrum based on the experimental flux values might be at disposal of the GEANT4 simulations for the space applications by using a probability grid. As exemplified in the present study, the experimental proton spectra together with the probability grid might be utilized to assess the shielding performance of the envisaged materials via the GEANT4 simulations.
\label{Conclusion}
\bibliographystyle{ieeetr}
\bibliography{Protonbiblio} 
\end{document}